\documentclass[prc,twocolumn,superscriptaddress,showpacs,amssymb,amsmath,amsfonts,aps]{revtex4}
\usepackage{graphicx}
\usepackage{dcolumn}
\begin{document}
\title{Measurement of the Polarized Structure Function $\sigma_{LT^\prime}$ \\
for Pion Electroproduction in the Roper Resonance Region} 

\newcommand*{\mpaa}{|M_{1+}|^2}
\newcommand*{\mpbb}{|E_{1+}|^2}
\newcommand*{\mpcc}{|S_{1+}|^2}
\newcommand*{\mpdd}{|M_{1-}|^2}
\newcommand*{\mpee}{|E_{0+}|^2}
\newcommand*{\mpff}{|S_{0+}|^2}
\newcommand*{\mpgg}{|S_{1-}|^2}
\newcommand*{\mpa}{M_{1+}}
\newcommand*{\mpb}{E_{1+}}
\newcommand*{\mpc}{S_{1+}}
\newcommand*{\mpd}{M_{1-}}
\newcommand*{\mpe}{E_{0+}}
\newcommand*{\mpf}{S_{0+}}
\newcommand*{\mpg}{S_{1-}}

\newcommand*{\UCONN }{ University of Connecticut, Storrs, Connecticut 06269} 
\affiliation{\UCONN } 

\newcommand*{\VIRGINIA }{ University of Virginia, Charlottesville, Virginia 22901} 
\affiliation{\VIRGINIA } 
 
\newcommand*{\YEREVAN }{ Yerevan Physics Institute, 375036 Yerevan, Armenia} 
\affiliation{\YEREVAN } 

\newcommand*{\JLAB }{ Thomas Jefferson National Accelerator Facility, Newport News, Virginia 23606} 
\affiliation{\JLAB } 
\newcommand*{\ASU}{Arizona State University, Tempe, Arizona 85287-1504}
\affiliation{\ASU}
\newcommand*{\UCLA}{University of California at Los Angeles, Los Angeles, California  90095-1547}
\affiliation{\UCLA}
\newcommand*{\CMU}{Carnegie Mellon University, Pittsburgh, Pennsylvania 15213}
\affiliation{\CMU}
\newcommand*{\CUA}{Catholic University of America, Washington, D.C. 20064}
\affiliation{\CUA}
\newcommand*{\SACLAY}{CEA-Saclay, Service de Physique Nucl\'eaire, F91191 Gif-sur-Yvette, France}
\affiliation{\SACLAY}
\newcommand*{\CNU}{Christopher Newport University, Newport News, Virginia 23606}
\affiliation{\CNU}
\newcommand*{\DUKE}{Duke University, Durham, North Carolina 27708-0305}
\affiliation{\DUKE}
\newcommand*{\ECOSSEE}{Edinburgh University, Edinburgh EH9 3JZ, United Kingdom}
\affiliation{\ECOSSEE}
\newcommand*{\FIU}{Florida International University, Miami, Florida 33199}
\affiliation{\FIU}
\newcommand*{\FSU}{Florida State University, Tallahassee, Florida 32306}
\affiliation{\FSU}
\newcommand*{\GWU}{The George Washington University, Washington, DC 20052}
\affiliation{\GWU}
\newcommand*{\ECOSSEG}{University of Glasgow, Glasgow G12 8QQ, United Kingdom}
\affiliation{\ECOSSEG}
\newcommand*{\ISU}{Idaho State University, Pocatello, Idaho 83209}
\affiliation{\ISU}
\newcommand*{\INFNFR}{INFN, Laboratori Nazionali di Frascati, Frascati, Italy}
\affiliation{\INFNFR}
\newcommand*{\INFNGE}{INFN, Sezione di Genova, 16146 Genova, Italy}
\affiliation{\INFNGE}
\newcommand*{\ORSAY}{Institut de Physique Nucleaire ORSAY, Orsay, France}
\affiliation{\ORSAY}
\newcommand*{\ITEP}{Institute of Theoretical and Experimental Physics, Moscow, 117259, Russia}
\affiliation{\ITEP}
\newcommand*{\JMU}{James Madison University, Harrisonburg, Virginia 22807}
\affiliation{\JMU}
\newcommand*{\KYUNGPOOK}{Kyungpook National University, Daegu 702-701, South Korea}
\affiliation{\KYUNGPOOK}
\newcommand*{\MIT}{Massachusetts Institute of Technology, Cambridge, Massachusetts  02139-4307}
\affiliation{\MIT}
\newcommand*{\UMASS}{University of Massachusetts, Amherst, Massachusetts  01003}
\affiliation{\UMASS}
\newcommand*{\MOSCOW}{Moscow State University, Skobeltsyn Nuclear Physics Institute, 119899 Moscow, Russia}
\affiliation{\MOSCOW}
\newcommand*{\UNH}{University of New Hampshire, Durham, New Hampshire 03824-3568}
\affiliation{\UNH}
\newcommand*{\NSU}{Norfolk State University, Norfolk, Virginia 23504}
\affiliation{\NSU}
\newcommand*{\OHIOU}{Ohio University, Athens, Ohio  45701}
\affiliation{\OHIOU}
\newcommand*{\ODU}{Old Dominion University, Norfolk, Virginia 23529}
\affiliation{\ODU}
\newcommand*{\PITT}{University of Pittsburgh, Pittsburgh, Pennsylvania 15260}
\affiliation{\PITT}
\newcommand*{\RPI}{Rensselaer Polytechnic Institute, Troy, New York 12180-3590}
\affiliation{\RPI}
\newcommand*{\RICE}{Rice University, Houston, Texas 77005-1892}
\affiliation{\RICE}
\newcommand*{\URICH}{University of Richmond, Richmond, Virginia 23173}
\affiliation{\URICH}

\newcommand*{\SCAROLINA}{University of South Carolina, Columbia, South Carolina 29208}
\affiliation{\SCAROLINA}
\newcommand*{\UNIONC}{Union College, Schenectady, NY 12308}
\affiliation{\UNIONC}
\newcommand*{\VT}{Virginia Polytechnic Institute and State University, Blacksburg, Virginia   24061-0435}
\affiliation{\VT}
\newcommand*{\WM}{College of William and Mary, Williamsburg, Virginia 23187-8795}
\affiliation{\WM}
\newcommand*{\NOWOHIOU}{Ohio University, Athens, Ohio  45701}
\newcommand*{\NOWINDSTRA}{Systems Planning and Analysis, Alexandria, Virginia 22311}
\newcommand*{\NOWUNH}{University of New Hampshire, Durham, New Hampshire 03824-3568}
\newcommand*{\NOWCUA}{Catholic University of America, Washington, D.C. 20064}
\newcommand*{\NOWSCAROLINA}{University of South Carolina, Columbia, South Carolina 29208}
\newcommand*{\NOWNONE}{unknown, }
\newcommand*{\NOWUMASS}{University of Massachusetts, Amherst, Massachusetts  01003}
\newcommand*{\NOWMIT}{Massachusetts Institute of Technology, Cambridge, Massachusetts  02139-4307}
\newcommand*{\NOWGEISSEN}{Physikalisches Institut der Universitaet Giessen, 35392 Giessen, Germany}
\newcommand*{\NOWSAK}{Sakarya University, Sakarya, Turkey}

\author{K.~Joo}
     \affiliation{\UCONN}
\author{L.C.~Smith}
     \affiliation{\VIRGINIA}
\author{I.G.~Aznauryan}
     \affiliation{\YEREVAN}
\author{V.D.~Burkert}
     \affiliation{\JLAB}
\author{H.~Egiyan}
\altaffiliation[Current address:]{\NOWUNH}
\affiliation{\WM}
\affiliation{\JLAB}
\author{R.~Minehart}
\affiliation{\VIRGINIA}
\author {G.~Adams} 
\affiliation{\RPI}
\author {P.~Ambrozewicz} 
\affiliation{\FIU}
\author {E.~Anciant} 
\affiliation{\SACLAY}
\author {M.~Anghinolfi} 
\affiliation{\INFNGE}
% ...

\collaboration{The CLAS Collaboration}

\noaffiliation
\begin{abstract}
{The polarized longitudinal-transverse structure function $\sigma_{LT^\prime}$ measures
the interference between real and imaginary amplitudes in pion electroproduction and
can be used to probe the coupling between resonant and non-resonant processes. 
We report new measurements of $\sigma_{LT^\prime}$ in the $N(1440)\frac{1}{2}^+$ (Roper) 
resonance region at $Q^2=0.40$ and $0.65$~GeV$^2$ for both the $\pi^0 p$ and $\pi^+ n$
channels.  The experiment was performed at 
Jefferson Lab with the CEBAF Large Acceptance Spectrometer (CLAS) using 
longitudinally polarized electrons at a beam energy of 1.515 GeV.  Complete angular 
distributions were obtained and are compared to recent phenomenological models.
The $\sigma_{LT^\prime}(\pi^+ n)$ channel shows a large sensitivity to the Roper resonance 
multipoles $M_{1-}$ and $S_{1-}$ and provides new constraints on models of resonance
formation.}
\end{abstract}
\maketitle

The structure of the $J^P=1/2^+$ $N(1440)$ resonance continues to be a mystery 
more than 40 years after its discovery by Roper~\cite{roper64} in the $P_{11}$
$\pi N$ channel.  Attention has largely centered on the inability of the
standard constituent quark model to describe the basic properties of this
resonance, such as its mass and photocouplings, and their $Q^2$ evolution. 
Quark models utilizing a harmonic or linear confining potential predict a 
normal level ordering of radial and orbital nucleon excitations according 
to parity, which is violated by the unusually low Roper mass.    This has raised questions
about the mechanism for breaking SU(6) symmetry in resonances, and alternatives 
to non-relativistic models with color-spin interactions~\cite{isgur77} between massive 
quarks have appeared.  These include relativistic treatments~\cite{capstick86}, 
such as modeling the Roper on the light-cone~\cite{capstick95}, or as a hybrid 
baryon ($q^3g$) where the state is assumed to have a large gluonic 
component~\cite{li92}, or as a baryon with a small quark core and a 
large meson cloud~\cite{cano98}, or as a $N\sigma$ molecule~\cite{kre00},
or even as a member of the pentaquark octet~\cite{jaffe03}. Finally, recent
quenched lattice QCD calculations~\cite{mat05} have shown that the observed
level ordering of the Roper emerges only in the chiral limit of vanishing quark mass.

The large variety of quark models make very distinct predictions for the internal structure
of the Roper, which can be tested by measuring the $Q^2$ dependence of 
the transverse $A^p_{1/2}$ and scalar $S^p_{1/2}$ photocoupling amplitudes. 
For example, the three-quark ($q^3$) state is predicted to have a 
characteristically slow falloff of $A^p_{1/2}$ and $S^p_{1/2}$. 
On the other hand, for the hybrid ($q^3g$) state $A^p_{1/2}(Q^2)$ 
is predicted to be more similar to the rapid falloff of the
$N \rightarrow \Delta(1232)$ transition, while $S^p_{1/2}(Q^2)$=0. 
Accurate knowledge of the Roper transition form factors therefore has 
significant implications for models of nucleon structure and understanding 
of the confinement mechanism.  

Electromagnetic studies of the Roper resonance have up to now been 
limited by the Roper's large width ($\approx$ 350 MeV)~\cite{eidelman04} and 
small photoproduction cross section.  Additionally, many of the data used for
such studies involve the $\pi^0 p$ final state, although the $\pi^+ n$
channel is more favorable due to the larger sensitivity to I=1/2 states.
Partial wave analysis (PWA) fits of cross-section 
measurements are necessary to separate the weak Roper excitation multipoles 
from non-resonant backgrounds and the tails of adjacent resonances.  However 
the reliability of this separation cannot be verified except through analysis
of additional experimental observables.  In particular, the polarized structure 
function $\sigma_{LT^\prime}$ in pion electroproduction measures the 
imaginary part of the interference between longitudinal (L) and transverse (T) amplitudes:
\begin{equation}
Im(L^*T)=Re(L)Im(T)-Im(L)Re(T)
\end{equation}
which can provide a powerful constraint to PWA fits.

\begin{figure}[h]
\includegraphics[scale=0.38]{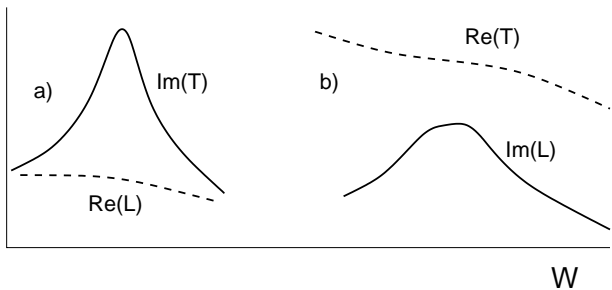}
\caption{Illustration of how $Im(L^*T)$ term can reveal different physics backgrounds. 
a) Weak background $Re(L)$ buried under strong resonance $Im(T)$. b) Weak resonance
$Im(L)$ buried under strong background $Re(T)$.  In each case interference through
Eq.~1 allows the stronger amplitude to amplify the weaker amplitude, making 
the latter experimentally accessible.} 
\label{fig:fig1}
\end{figure}

Recent measurements of $\sigma_{LT^\prime}$ in the $\Delta(1232)$ region~\cite{joo03,joo04}
showed a strong interference between the dominant $M_{1+}$ 
resonant multipole and largely real non-resonant backgrounds, as shown schematically 
in Figure~\ref{fig:fig1} (left).  In particular, the $\sigma_{LT^\prime}(\pi^+ n)$ channel
\cite{joo04} was well described by several phenomenological unitary models, indicating 
that the dominant $\it{t}$-channel pion pole and Born terms are under control. These 
Born contributions also determine the real parts of non-resonant multipoles in the Roper resonance region
and under the conditions illustrated in Figure~\ref{fig:fig1} (right) can greatly amplify
the small imaginary Roper resonant multipoles.

In this Brief Report we present the first measurements of $\sigma_{LT^\prime}$ obtained 
in the Roper resonance region using the $p(\vec e,e'\pi^+)n$ and $p(\vec e,e'p)\pi^0$ 
reactions. The data reported 
here span the invariant mass interval $W = 1.1 - 1.6$ GeV at $Q^2=0.40$ and $0.65$~GeV$^2$, 
and cover the full angular range in the $\pi N$ center-of-mass (c.m.). These
data were taken simultaneously with previous measurements in the $\Delta(1232)$ region
reported earlier~\cite{joo03,joo04}.

The experiment was performed at the Thomas Jefferson National 
Accelerator Facility (Jefferson Lab)
using a 1.515 GeV, 100\% duty-cycle beam of longitudinally 
polarized electrons incident on a liquid-hydrogen target. 
The electron polarization was determined by M{\o}ller polarimeter 
measurements to be 
$0.690\pm0.009$(stat.)$\pm0.013$(syst.). Scattered electrons and 
pions were detected in
the CLAS spectrometer~\cite{mec03}. Electron triggers were enabled
through a hardware coincidence of the gas \v{C}erenkov
counters and the lead-scintillator electromagnetic calorimeters.  
Particle identification
was accomplished using momentum reconstruction in the tracking system
and time-of-flight from the target to the scintillators. Software 
fiducial cuts were used to exclude regions of non-uniform detector response.
Kinematic corrections were applied to compensate for drift chamber
misalignments and uncertainties in the magnetic field.  
The $\pi N$ final state was identified using cuts on the missing hadronic mass. 
Target window backgrounds were suppressed with cuts on the reconstructed vertex. 

The single pion electroproduction cross section is given by:
\begin{equation}
\frac{d\,^4\sigma^h}{dQ^2 dW d\Omega^*_{\pi}} = J\,\Gamma_v\,\frac{d\,^2\sigma^h}{d\Omega^*_{\pi}},
\end{equation}
where $\Gamma_v$ is the virtual photon flux and the 
Jacobian $J = \partial(E',\cos\,\theta_e)/\partial(Q^2,W)$ relates the differential
volume element $dQ^2 dW$ of the binned data to the measured electron kinematics 
$dE'\,d\cos\,\theta_e$.  Here $d\,^2\sigma^h$ is the c.m.\ differential cross
section for $\gamma^* p \rightarrow \pi N$ with the electron beam helicity $h$. For an unpolarized target $d\,^2\sigma^h$ depends 
on the transverse $\epsilon$ and longitudinal $\epsilon_L$  polarization of the virtual 
photon through five structure functions: $\sigma_T,\sigma_L,\sigma_{TT}$, and the 
transverse-longitudinal interference terms $\sigma_{LT}$ and $\sigma_{LT^\prime}$:
\begin{eqnarray}
\frac{d\,^2\sigma^h}{d\Omega^*_{\pi}} &=& \frac{p^*_{\pi}}{k_{\gamma}^*} (\sigma_{0} +
h\sqrt{2\epsilon_L(1-\epsilon)}\,\sigma_{LT^\prime}\,\sin\,\theta^*_{\pi}\,\sin\,\phi^*_{\pi}),  \nonumber 
\\
\sigma_{0} &=& \sigma_T+\epsilon_L\sigma_L+\epsilon\,\sigma_{TT}\,\sin^2\theta^*_{\pi}\,\cos\,2\phi^*_{\pi} \nonumber \\
~&+&
\sqrt{2\epsilon_L(1+\epsilon)}\,\sigma_{LT}\,\sin\,\theta^*_{\pi}\,\cos\,\phi^*_{\pi},
\label{eq:str}
\end{eqnarray}
where $p_{\pi}^*$ and $\theta^*_{\pi}$ are the $\pi N$ c.m.\ momentum and polar angle, 
$\phi^*_{\pi}$ is the azimuthal rotation of the hadronic plane with respect to the electron
scattering plane, $\epsilon = (1+2|\vec{q}\,|^2\,\tan^2(\theta_e/2)/Q^2)^{-1}$, $\epsilon_L=(Q^2/|k^*|^2)\epsilon$, $|k^*|$ is the virtual photon c.m.\ momentum, and $k_{\gamma}^*$ is
the real photon equivalent energy.

Determination of $\sigma_{LT^\prime}$ was made through the beam spin 
asymmetry $A_{LT^\prime}$: 
\begin{eqnarray}
A_{LT^\prime} &=&\frac{d\,^2\sigma^+ - d\,^2\sigma^-}{d\,^2\sigma^+ +
d\,^2\sigma^-}  \\ 
&=&
\frac{\sqrt{2\epsilon_L(1-\epsilon)}\,\sigma_{LT^\prime}\,\sin\,\theta^*_{\pi}\,\sin\,\phi^*_{\pi}}{\sigma_{0}}.
\label{eq:altp}
\end{eqnarray}
The value of $A_{LT^\prime}$ 
was obtained for individual bins of $(Q^2,W,\cos\theta_{\pi}^*,\phi_{\pi}^*)$
by dividing the measured asymmetry $A_m$ by the
magnitude of the electron beam polarization $P_e$:
\begin{eqnarray}
A_{LT^\prime} &=& \frac{A_m}{P_e}  \\
A_m &=& \frac{N_\pi\,^+ - N_\pi\,^-}
{N_\pi\,^+ + N_\pi\,^-}, 
\label{eq:altp_m}
\end{eqnarray}
where $N_{\pi}^{\pm}$ is the number of livetime-corrected $\pi N$ events 
detected for each
electron beam helicity state, normalized to beam charge.
Radiative corrections were applied using the program 
recently developed by Afanasev {\it et al.} for exclusive pion 
electroproduction~\cite{aku02}. Corrections were 
also applied to compensate for cross section variations over the width
of each bin, using the MAID00 model described below.
Bin full-widths were $\Delta Q^2=0.15$~GeV$^2$, $\Delta W=0.4$~GeV, 
$\Delta\cos\theta_{\pi}^*=0.25$ and $\Delta\phi_{\pi}^*=45^0$.
Monte Carlo studies showed no significant helicity dependence to the
CLAS acceptance, therefore no acceptance corrections to $A_m$ were
applied.  Next the $A_{LT^\prime}$ distributions were multiplied 
by the unpolarized cross section $\sigma_{0}$,
using a parameterization of measurements of $\sigma_{0}$ made during the
same experiments~\cite{hov01,joo02}.  The structure function 
$\sigma_{LT^\prime}$ was then extracted using Eq.~\ref{eq:altp} by fitting 
the $\phi_{\pi}^*$ distributions corresponding to each $\cos\theta_{\pi}^*$ bin.
Systematic errors for $\sigma_{LT^\prime}$ were dominated by 
uncertainties in the determination of the electron beam polarization 
and the parameterization of $\sigma_{0}$. The systematic errors arising
from the other corrections to $A_m$ were negligible in comparison. 
Quadratic addition of the individual contributions yields a total relative 
systematic error of $< 6\%$ for all of our measured data points.

\begin{figure}[t]
\includegraphics[scale=0.47]{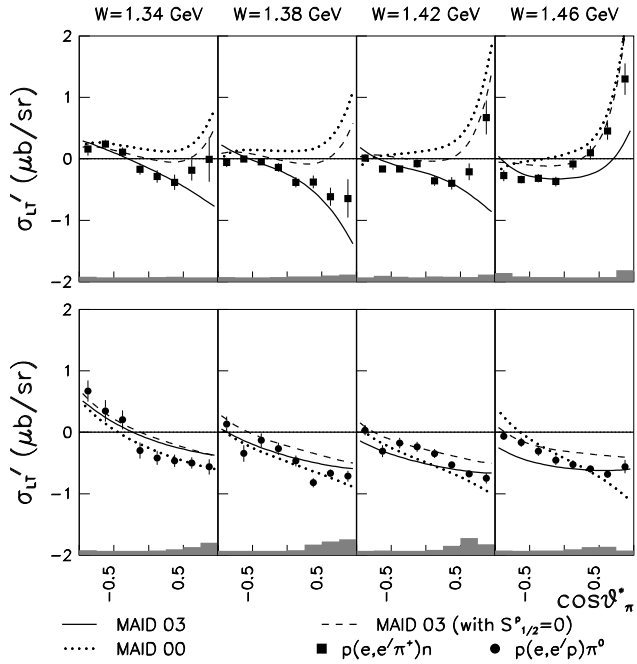}
\caption{CLAS measurements of $\sigma_{LT^\prime}$ versus $\cos\theta^*_\pi$ 
for the $\pi^+ n$ channel (top) and the $\pi^0 p$ channel (bottom) at
$Q^2$=0.40~GeV$^2$ and $W=1.34-1.46$~GeV. The curves show model predictions discussed
in the text. The shaded bars show estimated systematic errors.}
\label{fig:fig2}
\end{figure}

Figure~\ref{fig:fig2}  shows c.m.\ angular distributions 
of $\sigma_{LT^\prime}$ for different $W$ bins in the Roper
resonance region at $Q^2=0.40$~GeV$^2$. 
Our measurements for the $\pi^+ n$ channel (top) 
and the $\pi^0 p$ channel (bottom) are shown compared to the
unitary isobar model of Drechsel~{\it et al.} (MAID00 and MAID03)~\cite{dre99,tiat03},
a phenomenological parameterization of previous pion photo- and electroproduction
data.  MAID includes all well-established resonances parameterized using Breit-Wigner 
functions and with backgrounds calculated from Born diagrams and $\it{t}$-channel 
vector-meson exchange.  The model is unitarized according to
the $K$-matrix approach by incorporating the $\pi N$ scattering 
phase shifts~\cite{arnd04} into the background amplitudes and treating 
the rescattered pion as on-shell.  The MAID03 solution~\cite{tiat03} 
was fitted to recent $\pi^0$ electroproduction cross section data from Mainz, Bates, 
Bonn, and JLAB, while MAID00 estimated the transverse ($M_{1-}$) and longitudinal 
($S_{1-}$) Roper resonance photocouplings using older electroproduction data from the 1970s.

The structure function $\sigma_{LT^\prime}$ determines the imaginary part of 
bilinear products between longitudinal and transverse amplitudes and can be
expressed by the expansion:
\begin{eqnarray}
\sigma_{LT^\prime} & = & A + B P_1(\cos\theta^*_\pi) +  C P_2(\cos\theta^*_\pi),
\label{eq:str_legendre} 
\end{eqnarray}
with
\begin{eqnarray}
A &=& -{Im}(S_{0+}(M_{1-} - M_{1+} +3E_{1+})^* \nonumber \\
  &+& E_{0+}^*(S_{1-}-2S_{1+}) + ...) \\
B &=& -6{Im}(S_{1+}(M_{1-} - M_{1+} +3E_{1+})^* \nonumber \\
 &+& E_{1+}^*(S_{1-}-2S_{1+}) + ...) \\
C & = & -12{Im}((M_{2-} -  E_{2-})^*S_{1+} +  \nonumber \\
 &+& 2E_{1^+}^*S_{2^-} + ...), 
\label{eqn:rlt}
\end{eqnarray}
where $P_l(\cos\theta^*_\pi)$ is the $l^{th}$-order Legendre polynomial.
Sensitivity to the Roper multipoles $M_{1-},S_{1-}$ occurs mainly in the 
$A$ and $B$ Legendre coefficients, through interference with the electric and 
Coulomb dipole and quadrupole terms. The $t-$channel pion pole makes substantial 
contributions to $S_{0+},E_{1+}$ and $S_{1+}$ throughout the $\Delta(1232)$ and Roper 
regions, while the $s-$channel electric Born term saturates the $E_{0+}$
multipole.  For the $\pi^+$ channel, these multipoles are largely real
and significantly larger than for the $\pi^0$ channel.  As a result significant interference with the 
imaginary (resonant) parts of $M_{1-},S_{1-}$ is possible in the $\sigma_{LT^\prime}(\pi^+ n)$
observable.  This is demonstrated in Figure~\ref{fig:fig2}, where inclusion 
of a non-zero longitudinal coupling $S^p_{1/2}$ for the Roper drastically changes 
the shape of the MAID03 predicted $\sigma_{LT^\prime}(\pi^+ n)$ angular 
distributions (top), while the effect on $\sigma_{LT^\prime}(\pi^0 p)$ (bottom)
is much smaller.  

Our previous measurements~\cite{joo03,joo04} in the $\Delta(1232)$ 
resonance region were generally consistent with MAID03 for both
$\sigma_{LT^\prime}(\pi^0 p)$ and $\sigma_{LT^\prime}(\pi^+ n)$.
A pronounced forward peak was observed for $\sigma_{LT^\prime}(\pi^+ n)$,
which arose partly from the ${Im}(M_{1+}^*S_{1+})$ term,
but also received contributions from the interference of the $\Delta(1232)$
with the $\it{real}$ parts of $M_{1-},S_{1-}$.  The present CLAS measurement 
of $\sigma_{LT^\prime}(\pi^+ n)$ in Figure~\ref{fig:fig2}
clearly shows a supression of forward peaking similar to the MAID03 curve,
which in this $W$ region is due to a strong ${Im}(E_{1+}^*S_{1-})$ interference 
coming from the $\it{imaginary}$ part of the $S_{1-}$ Roper multipole 
in the $B$ Legendre coefficient. 

\begin{figure}[t]
\includegraphics[scale=0.47]{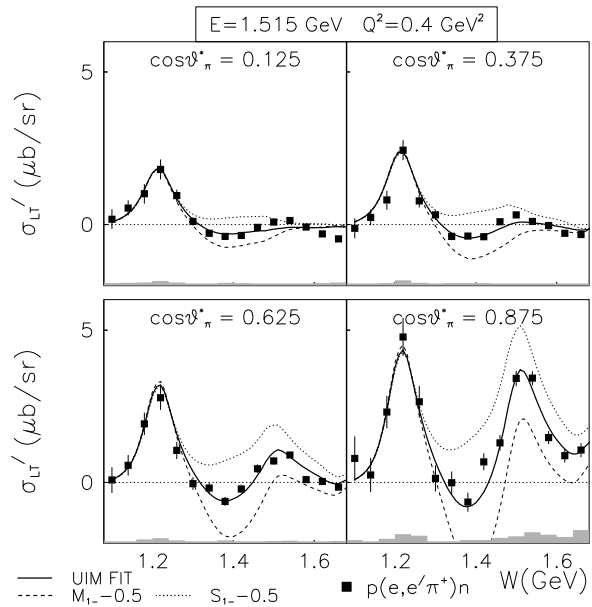}
\caption{CLAS measurements of $\sigma_{LT^\prime}$ versus $W$ (GeV)
for the $\pi^+ n$ channel extracted at $Q^2$=0.40~GeV$^2$ 
for different $\cos\theta^*_\pi$ points. The solid line shows the best fit
using the Unitary Isobar Model of Aznauryan~\cite{azn05}. 
The sensitivity of $\sigma_{LT^\prime}$ to the Roper resonance is
demonstrated by the dashed and dotted curves where the Roper contributions
to $M_{1-}$ and $S_{1-}$ are shifted by $-0.5~\mu b^{1/2}$. The shaded bars show estimated systematic errors.}
\label{fig:fig3}
\end{figure}

The significance of this interference is illustrated in 
Figure~\ref{fig:fig3}, which shows the $W$ dependence of 
$\sigma_{LT^\prime}(\pi^+ n)$ at $Q^2~=~0.4$ GeV$^2$ for 
different $\cos\theta_{\pi}^*$ bins, compared with the unitary 
isobar model (UIM) of Aznauryan~\cite{azn03,azn05}.  The resonant
photocoupling amplitudes in this model, which uses the same 
unitarization procedure as MAID, were determined from a global partial 
wave fit to all CLAS $\pi^0$ and $\pi^+$ electroproduction data 
(polarized and unpolarized) at $Q^2~=~0.4$ and 0.65 GeV$^2$, including
the data presented here.  The optimal fit reported in \cite{azn05} required
a large longitudinal photocoupling for the Roper, and a transverse
coupling near zero. Figure~\ref{fig:fig3} 
shows the UIM fit from \cite{azn05} after shifting the resonant part of 
each Roper multipole $M_{1-}$ and $S_{1-}$ by $-0.5~\mu b^{1/2}$, leaving 
the other at the fitted value.  This shift was comparable to the final fitted value of $S_{1-}$. 
It clearly shows that the sensitivity is larger in the $W$
region where the imaginary part of the Roper multipoles is nonzero, 
and maximized in the foward direction due to the interference 
through the pion pole term. 

In summary, we report new experimental measurements of the polarized structure 
function $\sigma_{LT^\prime}$ which show a large sensitivity to the Roper 
amplitudes in the $\pi^+ n$ channel through their interference with non-resonant 
backgrounds.  This is due to a combined effect of the isospin enhancement of 
the $\pi^+ n$ channel for $I=1/2$ resonances and the dominance of the $t-$channel 
pion pole term in the multipoles which interferes with the imaginary part of the 
Roper multipoles $M_{1-}$ and $S_{1-}$.  These data, in combination with other 
imaginary responses such as those extracted from recoil polarization experiments 
\cite{kel05}, will permit the most reliable determination of the resonant 
Roper photocoupling amplitudes.  This information can hopefully inspire new calculations
of the Roper transition form factor using modern hadronic models and lattice QCD.

We acknowledge the efforts of the staff of the Accelerator and Physics Divisions at 
the Thomas Jefferson National Accelerator Facility in their support of this experiment.  
This work was supported in part by the U.S. Department 
of Energy and National Science Foundation, an Emmy Noether Grant from the 
Deutsche Forschungsgemeinschaft, the French Commissariat \`a l'Energie 
Atomique, the Italian Istituto Nazionale di Fisica Nucleare, and the Korea Research 
Foundation. The Southeastern Universities Research Association (SURA) 
operates the Thomas Jefferson National Accelerator Facility for the United States Department 
of Energy under contract DE-AC05-84ER40150.

\end{document}